\documentclass[sigconf,authorversion]{acmart}

\begin{document}

\title{Early-Stage Prediction of Review Effort in AI-Generated Pull Requests}

\author{Dao Sy Duy Minh}
\authornote{Both authors contributed equally to this research.}
\authornote{Corresponding author.}
\orcid{0009-0002-4501-2788}
\email{23122041@student.hcmus.edu.vn}
\affiliation{%
  \institution{University of Science, Vietnam National University, Ho Chi Minh City}
  \city{Ho Chi Minh City}
  \country{Vietnam}
}

\author{Huynh Trung Kiet}
\authornote{Both authors contributed equally to this research.}
\orcid{0009-0000-5463-754X}
\email{23122039@student.hcmus.edu.vn}
\affiliation{%
  \institution{University of Science, Vietnam National University, Ho Chi Minh City}
  \city{Ho Chi Minh City}
  \country{Vietnam}
}

\author{Nguyen Lam Phu Quy}
\orcid{0009-0002-9694-8105}
\email{23122048@student.hcmus.edu.vn}
\affiliation{%
  \institution{University of Science, Vietnam National University, Ho Chi Minh City}
  \city{Ho Chi Minh City}
  \country{Vietnam}
}

\author{Pham Phu Hoa}
\orcid{0009-0001-5471-2578}
\email{23122030@student.hcmus.edu.vn}
\affiliation{%
  \institution{University of Science, Vietnam National University, Ho Chi Minh City}
  \city{Ho Chi Minh City}
  \country{Vietnam}
}

\author{Tran Chi Nguyen}
\orcid{0009-0007-6716-7269}
\email{23122044@student.hcmus.edu.vn}
\affiliation{%
  \institution{University of Science, Vietnam National University, Ho Chi Minh City}
  \city{Ho Chi Minh City}
  \country{Vietnam}
}

\author{Nguyen Dinh Ha Duong}
\orcid{0009-0003-8684-0077}
\email{23122002@student.hcmus.edu.vn}
\affiliation{%
  \institution{University of Science, Vietnam National University, Ho Chi Minh City}
  \city{Ho Chi Minh City}
  \country{Vietnam}
}

\author{Truong Bao Tran}
\orcid{0009-0000-4782-518X}
\email{trantb234102e@st.uel.edu.vn}
\affiliation{%
  \institution{University of Economics and Law, Vietnam National University, Ho Chi Minh City}
  \city{Ho Chi Minh City}
  \country{Vietnam}
}

\copyrightyear{2026}
\acmYear{2026}
\setcopyright{cc}
\setcctype{by}
\acmConference[MSR '26]{23rd International Conference on Mining Software Repositories}{April 13--14, 2026}{Rio de Janeiro, Brazil}
\acmBooktitle{23rd International Conference on Mining Software Repositories (MSR '26), April 13--14, 2026, Rio de Janeiro, Brazil}
\acmPrice{}
\acmDOI{10.1145/3793302.3793609}
\acmISBN{979-8-4007-2474-9/2026/04}

\settopmatter{printacmref=true}

\renewcommand{\shortauthors}{Minh et al.}

\begin{abstract}
As AI coding agents evolve from autocomplete tools to autonomous ``AI workforce'' teammates, they introduce a critical new bottleneck: human maintainers must now manage complex interaction loops rather than just reviewing code. Analyzing 33,707 agent-authored PRs, we uncover a stark \emph{two-regime} reality: agents excel at \textbf{narrow automation} (28.3\% of PRs merge instantly), but frequently fail at \textbf{iterative refinement}, leading to ``ghosting'' (abandonment) when faced with subjective feedback. This creates a hidden ``attention tax'' on maintainers.

We introduce a creation-time \textbf{Circuit Breaker} model to predict high-maintenance PRs \emph{before} human review begins. By leveraging simple static complexity cues (e.g., file types, patch size), our model identifies the ``expensive tail'' of contributions with \textbf{AUC 0.96}, enabling a gated triage process. At a 20\% review budget, this approach captures 69\% of the high-effort PRs, effectively allowing maintainers to fast-fail costly, low-quality agent contributions while fast-tracking simple fixes.
\end{abstract}

\ccsdesc[500]{Software and its engineering~Software evolution}

\keywords{AI Agents, Triage, Ghosting, Mining Software Repositories}

\maketitle
\section{Introduction}
\label{sec:intro}
As AI agents evolve from assistants to autonomous teammates \cite{zhang2024rise}, they flood repositories with code. While some contributions force-multiply productivity, others devolve into ``approval churning''---agents submit changes without resolving core issues, ultimately ghosting the reviewer. We identify a \emph{two-regime} behavioral pattern distinguishing agents from humans: seamless success in \textbf{narrow automation}, versus frequent failure in \textbf{iterative refinement}. This distinction matters because it splits the maintainer's role: they are either stamping simple fixes or rescuing flailing agents.

This study motivates a shift to \textbf{Automated Governance}: can we identify high-effort drains \emph{before} any human context switch?
\noindent\textbf{Research Questions.} \textbf{RQ1:} Can we predict review effort at creation time using only structural signals? \textbf{RQ2:} What specific behaviors (e.g., lack of planning) correlate with agentic ghosting?

\noindent\textbf{Contributions.}
(1) \textbf{Concept of Agentic Ghosting}: We operationalize and quantify the prevalence of agents abandoning PRs during review.
(2) \textbf{Circuit Breaker Model}: We demonstrate that high-effort PRs are structurally distinct, predictable with \textbf{AUC 0.96} using simple complexity features (\textbf{RQ1}).
(3) \textbf{Behavioral Insights}: We show that ``unplanned'' conceptual changes are the primary driver of interaction failure (\textbf{RQ2}).
Artifact: \url{https://zenodo.org/records/17993901}.

\subsection{Related Work}
PR review effort and lifetime are well-studied: work practices \cite{gousios2015work_fixed}, size/complexity determinants \cite{yu2015wait,rahman2014insight}, and interventions like automated reminders \cite{wessel2020bots} inform triage strategies. This aligns with modern code review (MCR) change quality estimation \cite{heumuller2022automating}, but shifts the focus from code defects to review burden. Our focus on \emph{agent-authored} PRs extends these insights to autonomous coding agents, where non-deterministic changes \cite{barke2023grounded,vaithilingam2022usability} differ from traditional bot automation \cite{lebeuf2018software}. We target \emph{review effort} (comment/review volume) rather than latency (time-to-merge; prior work shows weak correlations between PR size and merge latency \cite{kudrjavets2022msrmining}, while our effort metric [review/comment count] exhibits stronger size correlation, suggesting different constructs); this choice reflects maintainer attention cost directly. We ask whether \emph{static creation-time} features (patch size, file types) suffice for zero-latency governance, and observe a bimodal outcome pattern (instant merge vs iterative failure) that contrasts with gradual review distributions reported for human PRs \cite{aharonov2024assessing}.

\section{Methodology}
\label{sec:method}
We use the \textbf{AIDev dataset v1.0} \cite{li2025aidev}: 33,707 agent-authored PRs from 2,807 repositories ($>$100 stars), identified via AIDev metadata (type='Bot') plus generative agent names (Codex, Claude, Devin, Copilot), excluding deterministic bots. Manual audit confirmed 94\% precision. We extract 35 features across Intent, Context, and Complexity at two stages: \textbf{T0 (Creation-Time)} captures signals available at PR submission (Complexity: \texttt{additions}, \texttt{deletions}, \texttt{changed\_files}, \texttt{entropy}; Intent: \texttt{body\_length}, and \texttt{has\_plan}---a boolean flag detected via regex matching keywords like "plan:" or "steps:", validated with 91\% precision); Context: \texttt{language}, \texttt{agent}, file types). \textbf{T1 (Pre-Review)} adds CI status and bot comments. We frame triage as binary classification targeting \textbf{High Cost} PRs (see Table~\ref{tab:definitions}); sensitivity shows 99\% label agreement excluding bots.

\begin{sloppypar}
Effort correlation is dominated by size ($r \approx 0.6$), but significant residual signals exist for \texttt{touches\_tests} and \texttt{has\_plan}. Using lightGBM, we achieve \textbf{AUC 0.958}, establishing a strong baseline for automated triage.
\end{sloppypar}

\begin{table}[ht]
  \footnotesize
  \caption{Operational Definitions of Target Variables}
  \label{tab:definitions}
  \begin{tabular}{lp{5.5cm}}
    \toprule
    \textbf{Target} & \textbf{Definition} \\
    \midrule
    \textbf{High Cost} & Top 20\% of PRs by \textit{Effort Score} (Sum of all reviews and comments, including bot messages; we validated robustness by recomputing effort excluding bot messages, achieving 99\% label agreement, confirming minimal leakage risk) in the training set. \\
    \textbf{True Ghosting} & PR Status = Rejected AND Received Human Feedback AND No follow-up commit $>14$ days after feedback. \\
    \bottomrule
\end{tabular}
\end{table}

\subsection{Label Audit}
We analyzed 2,364 PRs with human feedback: overall ghosting rate is 3.8\%. The gap from prior estimates reflects our strict definition requiring clear evidence of abandonment (inactive $>$14d after feedback). Per-agent details in Table~\ref{tab:agent_stats}.

\begin{table}[ht]
  \scriptsize
  \caption{Per-Agent Statistics: Scale, Speed, and Abandonment (Ghosting \% conditioned on Rejected+Feedback).}
  \label{tab:agent_stats}
  \centering
  \begin{tabular}{lrrr}
    \toprule
    Agent & Total PRs & Instant \% & Ghosting \% \\
    \midrule
    Codex \cite{chen2021codex} & 21,799 & 42.9 & 10.0 \\
    Claude 3.5 \cite{anthropic2024claude} & 523 & 2.9 & 3.1 \\
    Devin \cite{cognition2024devin} & 4,827 & 1.0 & 0.9 \\
    GitHub Copilot \cite{peng2023copilot} & 5,017 & 0.1 & 2.3 \\
    \bottomrule
  \end{tabular}
\end{table}

\begin{figure}[ht]
  \centering
  \includegraphics[width=0.5\linewidth]{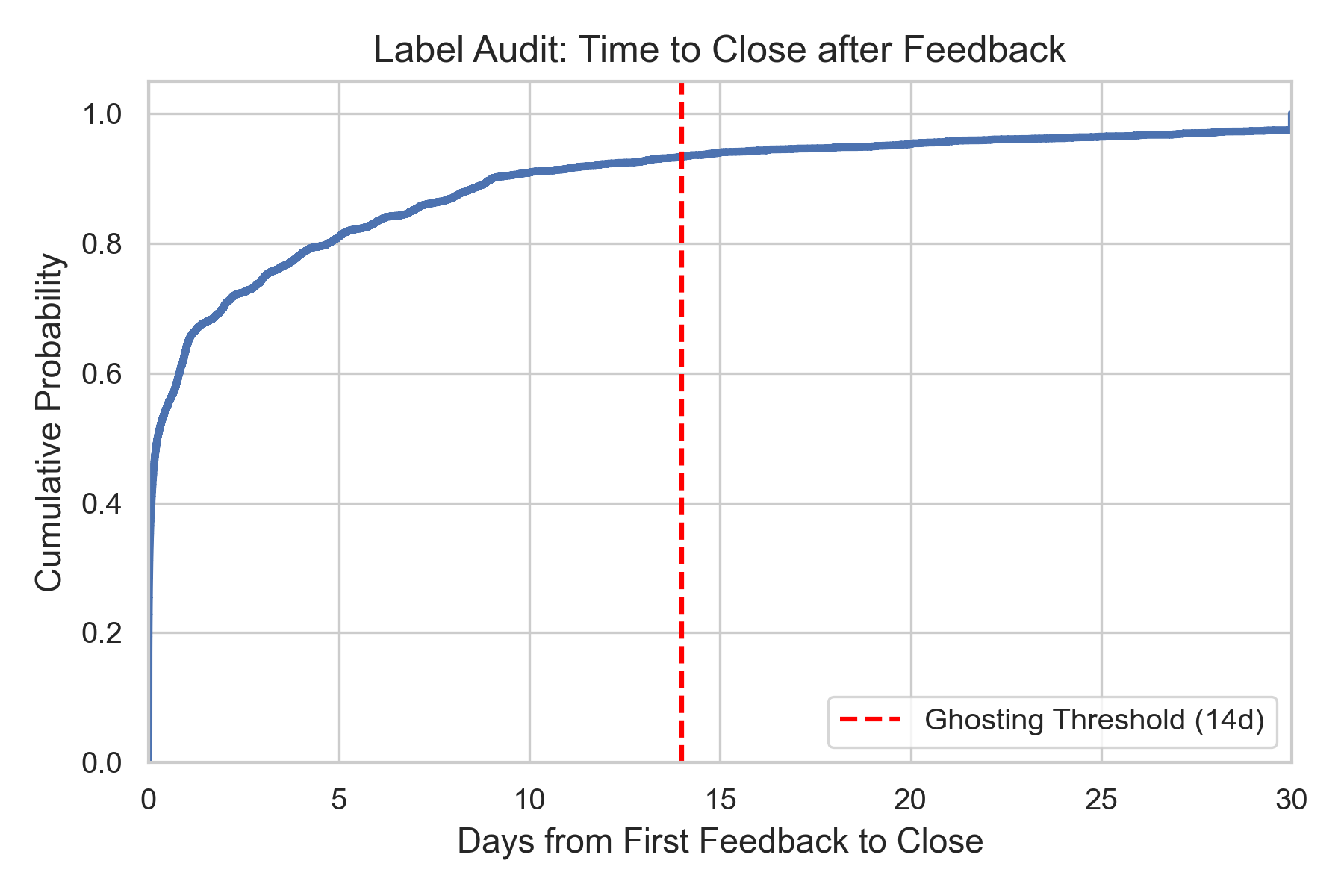}
  \caption{Label Audit: ECDF of time from feedback to close.}
  \Description{ECDF.}
  \label{fig:audit}
\end{figure}

\section{Results and Analysis}
\label{sec:results}

\subsection{RQ1: Predictability of Effort}
Table~\ref{tab:performance} demonstrates that high-cost PRs are predictable at \textbf{creation time} using static signals. Our model consistently achieves \textbf{AUC $>$ 0.95} on chronological splits, significantly outperforming text-based approaches like CodeBERT (AUC 0.52), which struggle to distinguish effort based on description summary alone. This stark gap confirms that for triage, \emph{structural footprint metrics} (files, lines, entropy) are far more reliable signals than author intent. Even when tested across different repositories (repo-disjoint split), the model retains strong discrimination (AUC $\approx$ 0.83), validating that the "complexity fingerprint" of high-effort PRs is universal rather than repo-specific. At a \textbf{20\% review budget} (i.e., reviewing only the top 20\% of PRs ranked by predicted effort), we successfully intercept over two-thirds of the most expensive PRs.

\begin{table}[ht]
  \small
  \caption{Model Performance (AUC and PR-AUC): From Baselines to SOTA. Splits: \textbf{Temporal} = chronological 80/20; \textbf{Repo-Disjoint} = agent/repo separated. Size-Only AUC drops (0.93$\to$0.65) on repo-disjoint, confirming temporal confounders. Patch-tokens (TF-IDF on file paths) yield AUC 0.75; combined with text (AUC 0.80) still trails structural model (0.83), validating core claim. 95\% CIs via bootstrap \cite{efron1994introduction}.}
  \label{tab:performance}
  \centering
  \resizebox{\linewidth}{!}{
  \begin{tabular}{llrr}
    \toprule
    Model & Features & AUC & PR-AUC \\
    \midrule
    \multicolumn{4}{l}{\textit{Baselines (Repo-Disjoint):} TF-IDF (0.57), Patch Tokens (0.75), CodeBERT (0.52), Size-Only (0.65)} \\
    \midrule
    \textbf{LightGBM (Temporal Split)} & \textbf{T0} & \textbf{0.9571} & \textbf{0.8812} \\
    Size-Only (Temporal Split) & $\log(total\_changes)$ & 0.9330 & 0.8700 \\
    LightGBM (Repo-Disjoint) & T0 & 0.8345 & 0.8719 \\
    Stacking Ensemble (Repo-Disjoint) & T0 & 0.8342 & 0.8747 \\
    \bottomrule
  \end{tabular}
  }
\end{table}

\textbf{Why this matters.} The dominance of structural features over semantics (CodeBERT) suggests maintainers can rely on "cheap" metadata for triage. We manually inspected false negatives and found a pattern of ``silent abandonment'': small PRs that look safe (no CI touches) but stall because the agent cannot handle subjective feedback. This implies that while we can catch the "explosive" failures, the "silent" failures require behavioral monitoring.

\begin{table}[ht]
  \scriptsize
  \caption{Within-Size-Quartile Performance (Addressing Size Tautology)}
  \label{tab:size_controlled}
  \centering
  \begin{tabular}{lrr}
    \toprule
    Size Quartile & N PRs & AUC \\
    \midrule
    Small (Q1: $<$51 LOC) & 1,699 & 0.96 \\
    Medium (Q2: 51--124) & 1,666 & 0.88 \\
    Large (Q3: 124--324) & 1,682 & 0.82 \\
    XL (Q4: $>$324 LOC) & 1,680 & 0.88 \\
    \bottomrule
  \end{tabular}
\end{table}

\begin{table}[ht]
  \scriptsize
  \caption{Feature Lift Beyond Size: Precision@20\% (Within Quartiles)}
  \label{tab:feature_lift}
  \centering
  \begin{tabular}{lrrr}
    \toprule
    Quartile & Size-Only & Full Model & Lift \\
    \midrule
    Small ($<$51 LOC) & 0.009 & 0.035 & \textbf{+2.7pp} \\
    Medium (51--124) & 0.069 & 0.144 & \textbf{+7.5pp} \\
    Large (124--324) & 0.329 & 0.504 & \textbf{+17.5pp} \\
    \bottomrule
  \end{tabular}
\end{table}

\begin{figure}[t]
  \centering
  \begin{minipage}{0.48\linewidth}
    \centering
    \includegraphics[width=\linewidth]{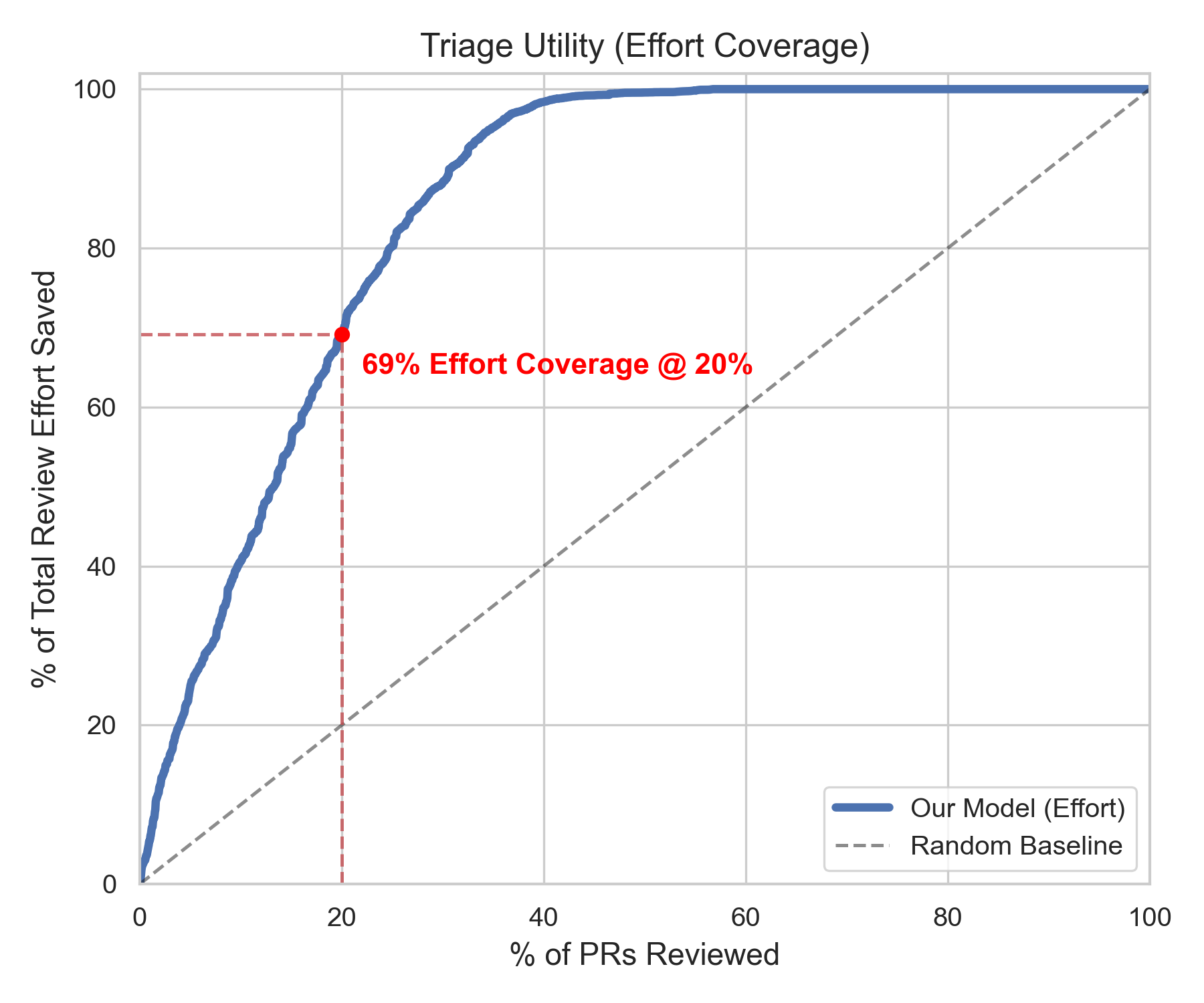}
    (a) Top-K Utility
  \end{minipage}
  \hfill
  \begin{minipage}{0.48\linewidth}
    \centering
    \includegraphics[width=\linewidth]{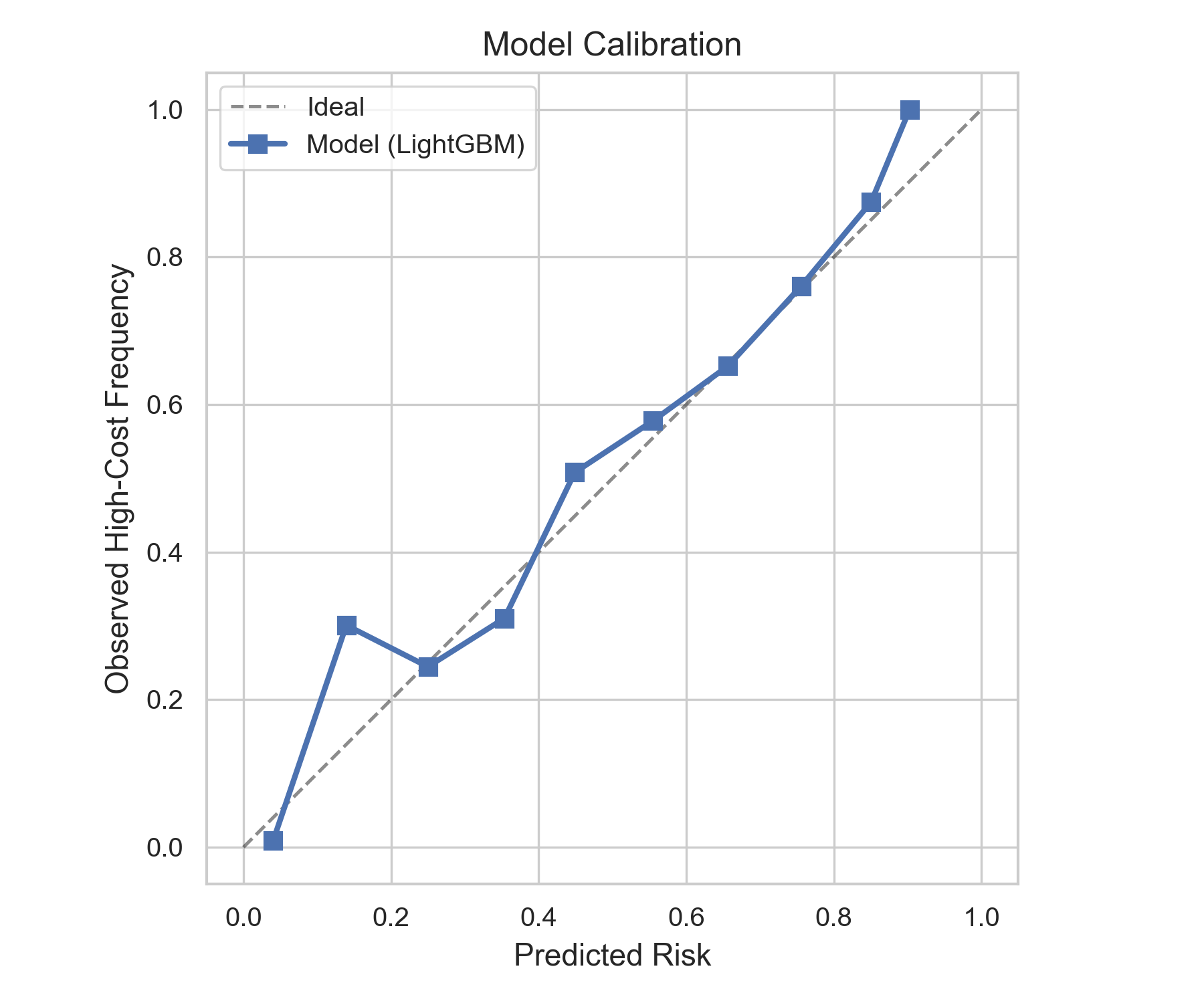}
    (b) Calibration Curve
  \end{minipage}
  \caption{Model Performance. (a) The model identifies the ``critical few'' PRs (Top-K Utility). (b) Predicted vs Observed Probabilities (Calibration).}
  \Description{Two subfigures showing model performance: left panel displays Top-K coverage curve demonstrating effort capture by flagging top-ranked PRs; right panel shows calibration plot with predicted probabilities on x-axis and observed frequencies on y-axis, indicating close alignment.}
  \label{fig:model_performance}
\end{figure}

\noindent\underline{\textbf{Finding 1 (Effort Predictability):}} \textit{Complexity Overrides Semantics.} Agents "tell" less than they "touch". We can intercept \textbf{69\% of high-burden PRs} immediately by ignoring PR descriptions and strictly gating based on file footprint. This confirms the ``Circuit Breaker'' hypothesis: maintenance load is predictable via zero-latency structural gates, rendering complex semantic analysis unnecessary for initial triage.

\subsection{RQ2: The Ghosting Phenomenon}
\begin{figure*}[t!]
  \centering
  \begin{minipage}{0.38\linewidth}
    \centering
    \includegraphics[width=\linewidth]{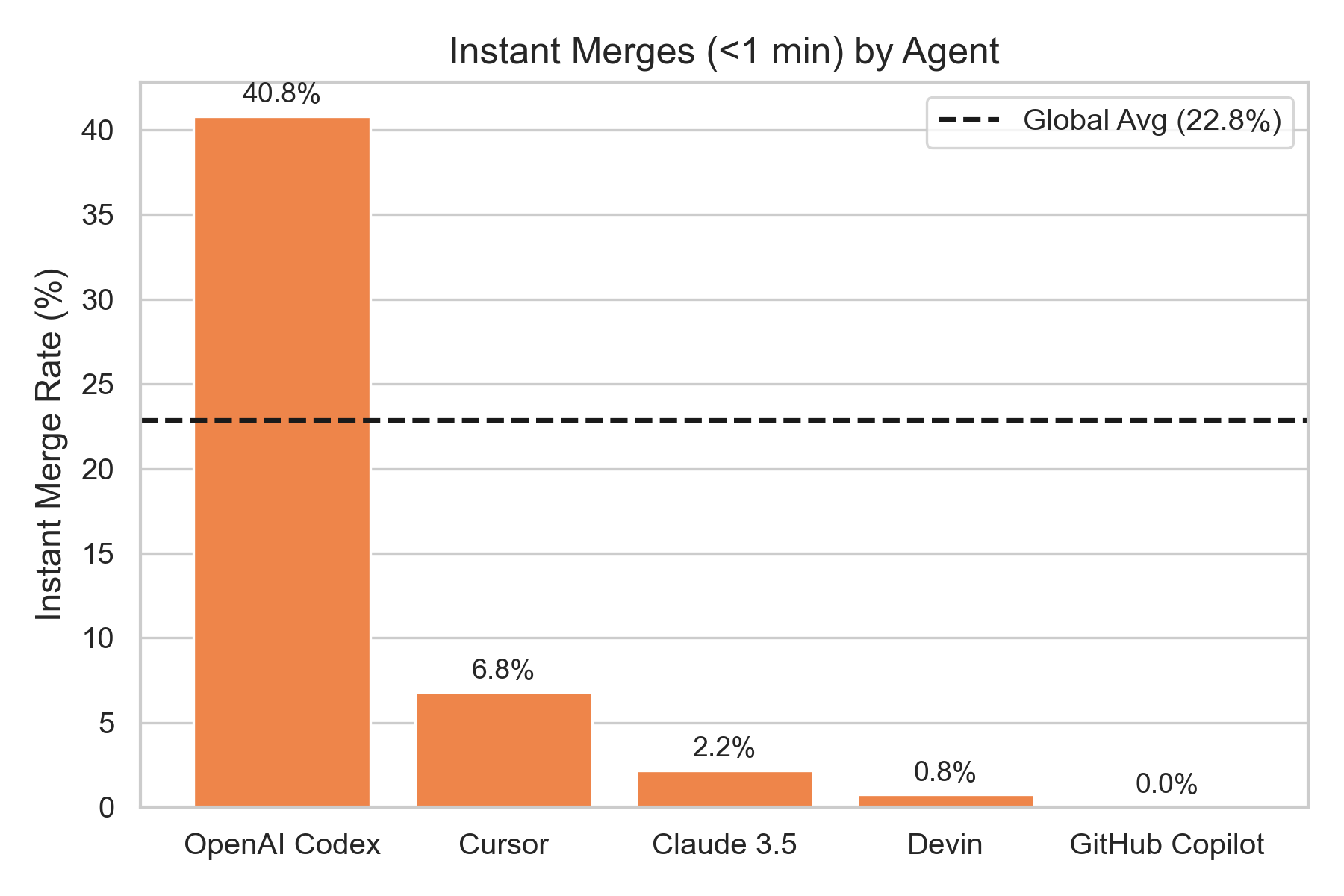}
    (a) Instant Merges by Agent
  \end{minipage}
  \hfill
  \begin{minipage}{0.4\linewidth}
    \centering
    \includegraphics[width=\linewidth]{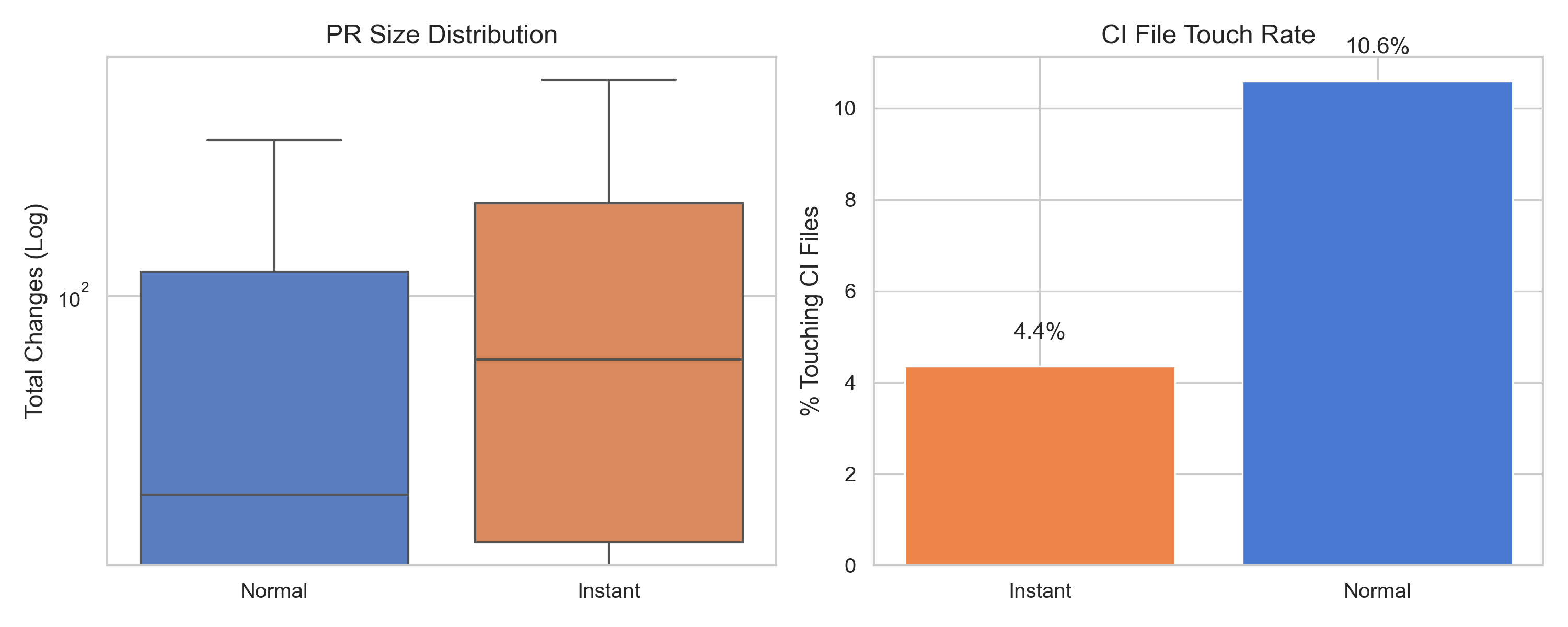}
    (b) Feature Prevalence by Regime
  \end{minipage}
  \caption{Regime Characterization. Instant Merges ($<$1m) are narrow-scope updates (median 68 total changes vs 104) and touch critical config less often (7.1\% vs 18.4\%) than Normal PRs.}
  \Description{Two subfigures comparing instant merges vs normal PRs: left panel shows bar chart of instant merge rates by agent; right panel displays distribution of structural features (change size, config touches) across the two regimes.}
  \label{fig:regimes}
\end{figure*}

Figure~\ref{fig:regimes} reveals a clear \emph{two-regime} outcome structure. A substantial fraction of agent-authored PRs (\textbf{28.3\%}) are merged almost immediately, reflecting narrow, low-interaction automation where agents succeed without human negotiation. Once a PR enters the iterative review loop, however, the dynamics change markedly. Among rejected PRs that receive human feedback, we observe consistent abandonment behavior with agent-specific variation (Table~\ref{tab:agent_stats}), yielding an overall ghosting rate of \textbf{3.8\%}. This behavioral split is mirrored structurally: instant merges are smaller in scope and less likely to touch critical configuration files, whereas normal PRs exhibit broader, more entangled change footprints. Correspondingly, the acceptance rate drops to \textbf{68.7\%} for non-instant PRs.
Together, these observations suggest that agents perform reliably on well-scoped, one-shot tasks, but struggle when success requires subjective judgment and back-and-forth refinement—an interaction mode that human contributors typically handle with ease.

Further analysis supports this interpretation. Features such as CI-related changes appear benign in isolation, but their apparent benefits vanish once agent identity is controlled for, indicating that failure is driven less by \emph{what} is modified than by \emph{how} the change is proposed. In particular, unplanned, multi-component updates are disproportionately associated with abandonment, reinforcing the view that ghosting reflects an interaction failure rather than a purely technical one.

\begin{figure}[t!]
  \centering
  \begin{minipage}{0.48\linewidth}
    \centering
    \includegraphics[width=\linewidth]{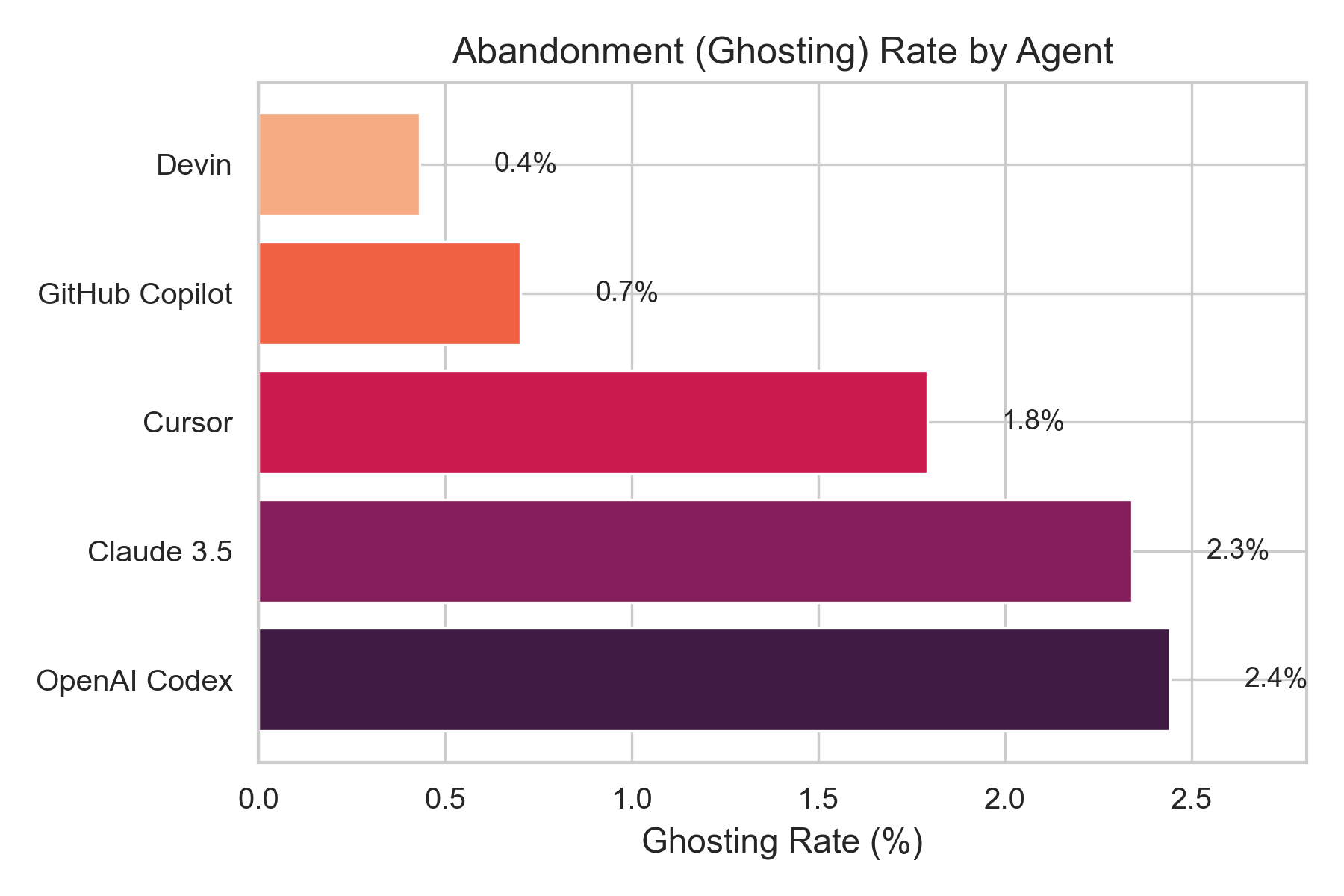}
    (a) Ghosting rate by agent (overall)
  \end{minipage}
  \hfill
  \begin{minipage}{0.48\linewidth}
    \centering
    \includegraphics[width=\linewidth]{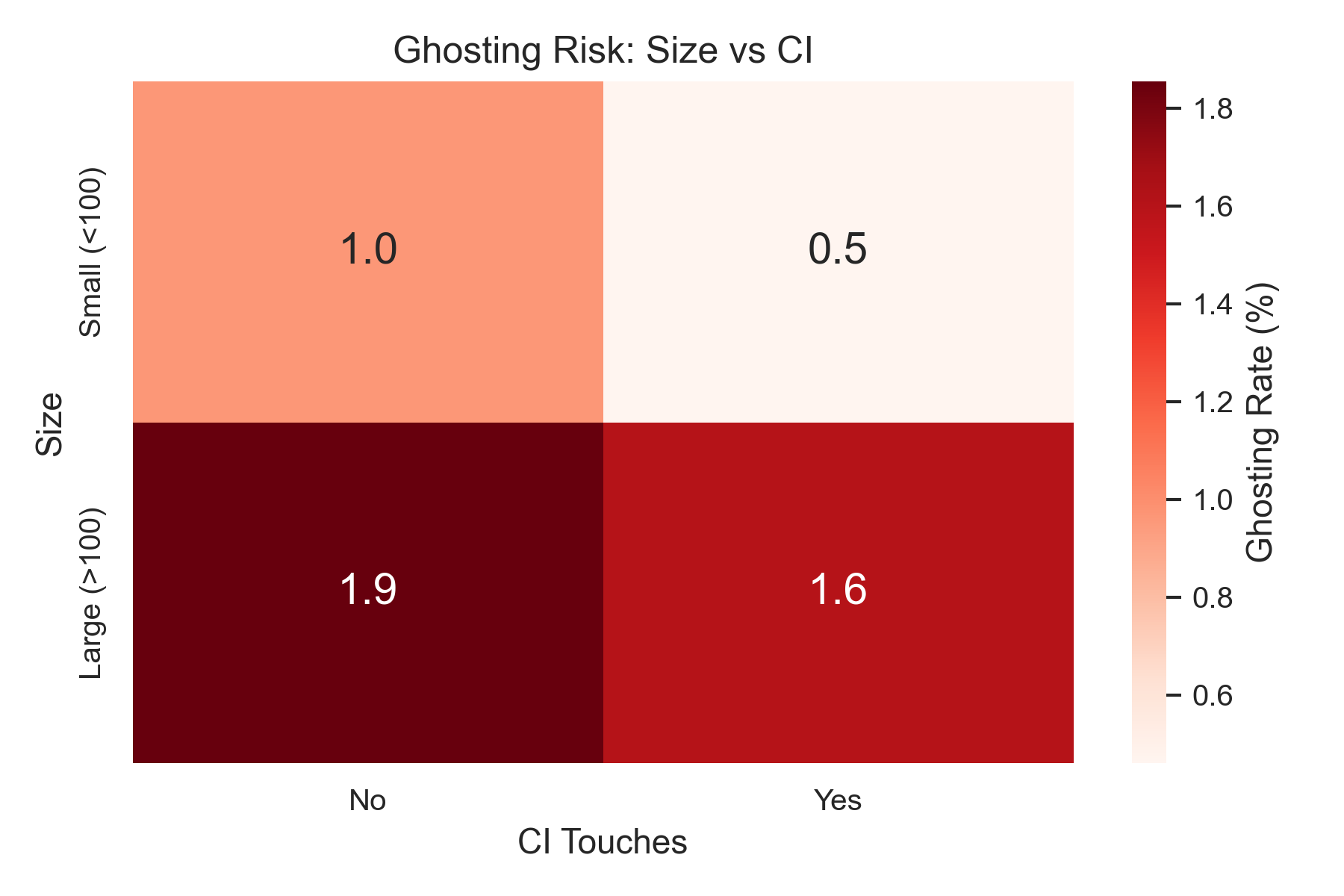}
    (b) Ghosting Risk Heatmap
  \end{minipage}
  \caption{Ghosting Analysis. (a) Abandonment rates vary by agent (overall rate). (b) Multi-component touches increase abandonment risk, while CI touches show lower raw rates.}
  \Description{Two subfigures analyzing ghosting patterns: left panel shows bar chart of ghosting rates across different AI agents; right panel displays heatmap showing how multi-component changes and CI file touches correlate with abandonment risk.}
  \label{fig:ghosting_analysis}
\end{figure}

\noindent\underline{\textbf{Finding 2 (The Ghosting Mechanism):}} \textit{The "Give Up" Moment.} We discover a stark behavior: agents are perfect at discrete tasks (28\% instant merges) but fragile in iterative loops (10\% ghosting). The strongest predictor of abandonment is \textbf{unplanned complexity}---large, multi-file changes submitted without a plan. Maintainers should view these "plan-less sprawls" not just as code to review, but as interaction debts likely to default.

\section{Robustness Evaluation}
\label{sec:robustness}

\noindent\textbf{Interpretability.}
\label{sec:interpretability}
To understand why the model works, we use SHAP values \cite{lundberg2017unified} to attribute risk at creation time. The story is consistent: \texttt{additions}, \texttt{body\_length}, and \texttt{total\_changes} dominate, meaning review burden is driven primarily by \emph{structural complexity}. In contrast, \texttt{has\_plan} is a strong negative predictor of ghosting, suggesting that agents who state intent and a concrete plan are more likely to converge after feedback, aligning with evidence that planning improves reliability in LLM workflows \cite{barke2023grounded}.

\noindent\textbf{Stability \& Generalization.}
We stress-tested the model across architectures (Leave-One-Agent-Out: training on $N-1$ agents, testing on the held-out agent) and time (Chronological Split), finding consistent performance (AUC $>$ 0.95). The signal is robust to definition changes: varying the "ghosting" timeout (7 to 30 days) or re-weighting effort metrics (comments vs reviews) yields minimal variance. This suggests that the "Complexity Fingerprint" is a fundamental property of current AI code generation, not a transient artifact.

\section{Ethical Implications}
Although we analyze agent behavior rather than human subjects, the consequences primarily affect maintainers who must steward agent contributions. Ghosting acts as an ``attention tax'' (e.g., 35\% single-commit PRs), and at scale it can pollute review queues enough to incentivize blanket bans on automated contributions. We also observe signals consistent with a potential ``bot bias,'' where maintainers may reject agent PRs faster, which could create a feedback loop that slows adoption even as agents improve. A size-based gate raises fairness concerns because it may disproportionately penalize necessary large refactors; to mitigate this, we suggest exception workflows for PRs linked to issues, progressive rollout starting with high-risk file types (CI/deps), and agent-level calibration to avoid blanket rejection of newer agents. Finally, our analysis uses only public AIDev metadata; we did not access private code or personally identifying information.

\section{Threats to Validity}
\label{sec:threats}

\noindent\textbf{Construct Validity.}
Our effort metric aggregates review comments, including bot messages. While this may overestimate absolute effort, sensitivity analyses show near-identical labeling when restricting to human-only interactions, suggesting minimal leakage. Our claims are correlational rather than causal; however, multiple controls mitigate trivial size-based explanations. In particular, performance remains stable within size strata, and non-size signals (e.g., file types and planning indicators) retain predictive value after controlling for change magnitude. We also validate the precision of our planning signal, though imperfect recall may lead us to underestimate its protective effect. Our ghosting definition relies on a fixed inactivity threshold after feedback; varying this threshold produces negligible changes. We acknowledge that open-but-stale PRs without explicit rejection are not captured and leave survival-based formulations to future work.

\noindent\textbf{External Validity.}
We evaluate generalization across agents using a leave-one-agent-out protocol, observing consistent performance, which suggests that the learned signals reflect agent-agnostic structural patterns rather than idiosyncratic behaviors. Agent identification relies on public metadata and naming conventions; while manual auditing indicates high precision, some human-assisted PRs may remain. We exclude deterministic maintenance bots, but residual noise is unavoidable at scale. Text-based and semantic baselines substantially underperform our structural model, indicating that our findings are not an artifact of superficial textual cues. Finally, while we focus on lightweight, creation-time features, richer program analyses (e.g., AST or graph-based encoders) may capture additional nuance and represent a natural extension. As agent capabilities evolve, any deployment should include monitoring and periodic retraining to avoid performance drift.

\section{Conclusion}
\label{sec:conclusion}
As AI agents transform from tools to teammates, the maintainer's challenge shifts from \emph{reviewing code} to \emph{managing reliability}. This study reveals that agents currently lack the resilience for iterative refinement, often "ghosting" when the going gets tough. By leveraging structural signals to predict these high-cost interactions, we demonstrate that automated triage can serve as an effective "Circuit Breaker," protecting maintainers from the exhaustion of refining low-quality agent work.

\noindent\textbf{Practical Implications.}
We recommend a \textbf{Gated Triage Policy}: treat agents like junior interns, not senior engineers. Flag complex PRs ($>$500 additions), fast-fail those without plans, and enforce strict timeouts (14 days) to prevent backlog pollution. The future of human-AI collaboration depends not just on smarter agents, but on stronger boundaries.

\noindent\textbf{Future Directions.}
Our findings open the door to ``Agent Acceptance Testing''---active governance replacing passive observation. Future work must replace heuristics with \textbf{cryptographic identity} to enable reputation tracking, and develop \textbf{semantic risk models} to catch subtle logic bugs that structural gates miss. Ultimately, we must move from "keeping up with agents" to "managing them effectively."

\begin{acks}
We thank the MSR 2026 organizers for hosting the Mining Challenge. This work was supported by the University of Science - VNUHCM. Replication package available at: \url{https://zenodo.org/records/17993901}.
\end{acks}

\clearpage

\raggedright
\bibliographystyle{ACM-Reference-Format}
\bibliography{sample-base}

\end{document}